\documentclass[aps,prb,
,amssymb]{revtex4} 
\usepackage{epsfig}

\begin{document}
\renewcommand{\thefootnote}{\fnsymbol{footnote}}
\renewcommand{\theequation}{\arabic{section}.\arabic{equation}}

\title{Global exploration of the energy landscape of solids on the
{\it ab initio} level}

\author{K. Doll, J.C. Sch{\"{o}}n and M. Jansen}

\affiliation{Max-Planck-Institute for Solid State Research,\\
Heisenbergstr. 1, D-70569 Stuttgart, Germany.
}
\date{\today}


\begin{abstract}
\noindent 
Predicting which crystalline modifications can be present in a
  chemical system
requires the global exploration of its energy landscape. Due to the large
  computational effort involved, in the past 
this search for sufficiently
stable minima has been performed employing a variety of empirical potentials
and cost functions
followed by a local optimization on the {\it ab initio} level. However, this
entails the risk of overlooking
important modifications that are not modeled accurately using empirical
potentials. In order to overcome
this critical limitation, we develop an approach to employ {\it ab initio} 
energy functions during the
global optimization phase of the structure prediction. As an example, we 
perform a global exploration
of the landscape of LiF on the {\it ab initio} level
and show that the relevant 
crystalline modifications are found during the search.
\end{abstract}

\maketitle

\section{Introduction}
A fundamental issue in solid state theory is the crystalline structure a
given chemical system exhibits in the solid
state\cite{Maddox88,Cohen89,Hawthorne90,Catlow90,Schoen96b,Jansen02b}. 
Why is a particular
periodic atomic configuration adopted, which among several modifications
is the preferred one at a particular temperature and pressure, and which
thermodynamically metastable but kinetically stable modifications
are possible in the first place? Answering these questions requires the
global exploration of the energy landscape of the chemical system
\cite{Schoen96b,Jansen02b,Schoen01}.
Every metastable modification of a solid compound corresponds to a locally 
ergodic region on the energy landscape\cite{Schoen96b,Schoen01}, 
i.e.\ a set of atomic 
configurations which exhibits the property that the equilibration 
time $\tau_{eq}$ of the chemical system within the region is much
smaller than the observational time scale $t_{obs}$, which in its
turn is much smaller than the time scale $\tau_{esc}$ on which the
system is expected to leave this region, 
$\tau_{eq} \ll t_{obs} \ll \tau_{esc}$. In particular at low
temperatures such locally ergodic regions constitute basins around
one or several local minima of the energy landscape, and the kinetic
stability of the corresponding compounds is controlled by the energetic
and entropic barriers surrounding the region\cite{Schoen01,Schoen03}. 
Thus, the first
step in the prediction of the possible structures in a chemical system
is the determination of the local minima on the energy landscape using
a global optimization algorithm. One should note that it is not sufficient
to obtain only the global minimum: all local minima that are surrounded by
sufficiently high energy barriers correspond to metastable modifications
that may be of interest regarding their physical and/or chemical properties
both in scientific and in technological applications
\cite{Schoen96b,Jansen02b}.

Since the beginning of the 1990's, methods for theoretical structure
determination and prediction employing global optimization techniques
have been developed
\cite{Liu90,Pannetier90,Freeman93,Schoen94,Boisen94,Schoen95,Bush95,Putz98a,Woodley99,Mellot00,Allan00,Mellot02,Winkler01,Oganov06,Pentin06a},
using e.g. simulated annealing \cite{Kirk83,Czerny85}, genetic algorithms
\cite{Holland75,Johnston04,Woodley07}, or the threshold algorithm\cite{Sibani93,Schoen96a}. 
Recently, the combination
of molecular dynamics with a history dependent potential was suggested
in the framework of the
metadynamics approach \cite{Laio2002}, in order to explore energy landscapes
and phase transitions, e.g. \cite{Martonak2005}. The general idea, i.e.\ starting at a local minimum and
exploring the neighborhood on the landscape bears some resemblance to the lid or
threshold algorithm\cite{Sibani93,Schoen96a}, but is different from typical global
optimization techniques such as simulated annealing, where long jumps on the energy
landscape are allowed (sometimes called 'basin hopping'). The metadynamics employs
molecular dynamics and 
focuses
on a set of a few relevant variables which are used to define excluded regions of the landscape
and to describe the
reaction mechanism, while in the threshold algorithm the energy surface is stochastically
sampled by a 
Monte-Carlo random walk below a sequence of energy thresholds, and the lid algorithm 
systematically explores the landscape below such energy lids by excluding all parts of phase
space that have already been visited.

Since a typical set of global optimization
runs involves millions or even billions of energy evaluations, a modular
approach has become standard\cite{Schoen96b,Woodley04}, where a global 
search on an empirical
energy / cost function landscape generates structure candidates, which are
subsequently locally optimized on full quantum mechanical level using
e.g.\ the Hartree-Fock approximation or density functional
theory\cite{Putz98a}.
Of course, this use of empirical potentials contains the risk that good
candidates are overlooked because they do not correspond to a
minimum (or only to a high-lying shallow one) on the empirical landscape, and there are many
chemical systems where no straightforward empirical energy function based on simple or
refined potentials or a crystal-chemically inspired cost function such as a bond-valence potential exist.
But even for those systems such as ionic compounds where supposedly good model potentials
have been constructed, the question to what degree the empirical energy landscape globally
agrees with the {\it ab initio} energy landcape has been debated since the inception of work on
structure prediction\cite{Schoen96b,Cancarevic06,Martonak06}. Clearly, a careful comparison between these
two energy landscapes  for a particular system should yield much insight into the foundations
of the current standard modular approach to structure prediction in solids.
Only now the computers are reaching the speed and ubiquity that will
allow us to perform the global optimization on the {\it ab initio} level, as
Oganov and co-workers\cite{Oganov06} have shown using a hybrid genetic
algorithm for this purpose.

However,
the step from model potentials to
{\it ab initio} calculations is absolutely
non-trivial, and requires careful adjustment of parameters to
use only a miminum of CPU time and thus keep
the calculations tractable. 
In this study, we investigate the energy landscape of the LiF-system
using stochastic simulated annealing. This system was
chosen as a test case, since we had studied the landscapes of the alkali
halides in earlier work\cite{Schoen95} using a Coulomb- plus
Lennard-Jones-potential. Thus, the most important structure candidates in the
system are known at the 
empirical potential level, and we can better judge both 
the success of the global
exploration on the {\it ab initio} level and the degree of agreement between the empirical and
{\it ab initio} energy landscapes than would have been possible when 
choosing
a not-yet-investigated chemical system. 

There are thus two major goals of this article:
Firstly, to show that
the {\it ab-initio} exploration of energy landscapes with a Monte Carlo random walk based technique
such as simulated annealing
is feasible. Secondly,
to investigate to what degree crucial features of the landscape such as the relevant minima 
on the level of the empirical
potentials and on the {\it ab initio} level are the same. Finally, we note that the {\it ab initio} energy
landscape is expected to be an appropriate choice for any system, whether ionic, covalent, or metallic.
Thus, being able to globally explore such an {\it ab initio} energy landscape will open the path to 
structure prediction in systems which can no longer be reasonably described with 
straightforward empirical potentials.

\section{Methods}
\label{meth}
\subsection {General approach.}
\label{gen-ap}
Our general approach to the determination of structure candidates has
been given in detail elsewhere\cite{Schoen96b,Schoen01,Schoen05}. To
summarize: First, the minima on the energy landscape are determined using
simulated annealing, possibly combined with a stochastic quench, as a global
optimization algorithm, where both atom positions and cell parameters are
freely varied. Next, the corresponding configurations are analyzed regarding
their symmetries using an algorithm to find symmetries \cite{Hundt99a} and
the space group
\cite{Hannemann98a} as implemented in the program KPLOT\cite{Hundt79},
followed by a comparison using an algorithm to compare cells\cite{Hundt06}, 
in order to
eliminate duplicate structures. Finally, the structure candidates are locally
optimized on the {\it ab initio} level using both a heuristic
algorithm\cite{Cancarevic04a,Schoen04b} and the energy minimization routines
included in the various {\it ab initio} codes. We always employ several {\it ab initio}
methods (Hartree-Fock and density functional theory), since this allows us to
compare the ranking of the candidates by energy and thus to gain some estimate
of their thermodynamic stability. This is crucial since, by definition, no
comparison of the predicted structures with experimental data is possible and
thus we cannot "tune" the parameters of the quantum mechanical methods to
reproduce the experiment. Furthermore, comparing the outcomes of the local
optimizations for different methods yields insights into the connectedness of
the candidates via low-lying saddles on the energy landscape. Finally, if
sufficient computational power is available, we can employ the lid
or threshold algorithm\cite{Sibani93,Schoen96a}, in order to quantitatively
study the energetic and entropic barriers on the landscape, which control the
kinetic stability of the metastable modifications of the solid compound.  

\subsection{ {\it Ab initio} calculations and global exploration: technical details.}
\label{abin}
For the {\it ab initio} energy calculations we employ the program 
CRYSTAL2006 \cite{CRYSTAL2006}. 
A set of preliminary tests was performed to optimize the efficiency of our approach
when applied to structure prediction on the {\it ab initio} level.
The most important parameters tested were:
the basis sets for the {\it ab initio} calculations;
parameters such as integral thresholds for the {\it ab initio} calculations;
the length of the simulated annealing run and of the subsequent quench run;
the move classes involved.

This preliminary step is actually crucial for the task of performing
{\it ab initio} explorations of energy landscapes: the energy calculations
are performed without the use of symmetry, since all possible structures must be accessible
during the random walk.
However, with the default parameters in CRYSTAL2006, a single
Hartree-Fock calculation for
an eight-atom simulation cell whose side length equals to the experimental lattice constant without symmetry (space group P1)
takes $\sim$ 13 minutes.
A typical run may consist of 100000 and more simulated annealing steps,
and thus the total CPU time for a single run would be on the scale of
10$^6$ minutes, i.e. roughly 2 years. For the exploration of a landscape,
dozens or even hundreds of such runs are necessary. In addition, it is
necessary to achieve convergence of the self-consistency cycles
in calculations which start from a random geometry.
This makes it obvious that a very
careful calibration of all parameters is necessary. One has
to exploit the fact that only a rough knowledge about the possible
local mimima is required in the first stage of the global optimization.
The final local optimization, based on
an optimization via analytical gradients, can be subsequently done with good parameters.

The initial tests resulted in the following choices:
The basis sets from reference \cite{Prencipe} were selected, with a slightly modified
fluorine basis set: 
slightly tighter $sp$
functions were chosen for the two outermost exponents  
(0.45 instead of 0.437, 0.2 instead of 0.147) to enhance the numerical
stability and the speed of the calculations.
As the global optimization only has to provide a rough information about the
energy landscape, it is not necessary to converge the solution very
accurately. The threshold for the convergence of the self-consistent field
(SCF) cycle was therefore reduced from $10^{-5}$ to $10^{-3} \ E_h$. 
Similarly,  the thresholds for
neglecting integrals were reduced from the default values
of $10^{-6}$, $10^{-6}$, $10^{-6}$, $10^{-6}$, $10^{-12}$ to 
$10^{-4}$, $10^{-4}$, $10^{-4}$, $10^{-4}$, $10^{-8}$. For the $\vec k$-point
sampling, a shrinking factor
of 2 was used. The
calculations were performed at the Hartree-Fock level. 
Note that since only the approximate positions of the
basins need to be found during the initial global exploration, the level of
theory does not play a crucial role.

The initial cell was cubic with a cell parameter of 7.07 \AA. 
4 lithium and 4 fluorine atoms were randomly placed in this cell.
No symmetry was used, i.e. the simulated annealing and quenching
was done in $P1$.
The probabilities of the individual moves were as follows:
moving individual atoms (70 \%; maximal step size was 5 pm), exchanging atoms (10 \%),
moving atoms with (10\%) and without (5\%) simultaneous change of the
unit cell, and the change of the origin (5 \%; this
move is important when the cell is subsequently truncated due to a change of the
cell vectors). For all the moves which change 
the cell parameters, the probability
of a suggested move to shorten the cell was set to 70\%, in order to speed up the
shrinking of the cell. To avoid atoms coming too close, a minimum
distance of the sum of the radii, multiplied by 0.8, was required.
The radii were determined by using the Mulliken charges and linearly
interpolating between the tabulated radii \cite{Emsley90} for the neutral atoms 
and the ions. With this choice of parameters, it turned out that only
a very short simulated annealing run was required (5000 steps,
with the starting and final temperatures corresponding to 1 eV and
0.9 eV, respectively), followed by a quench of $\sim$ 5000 steps. The reason why the 
simulated annealing part could be kept so short is probably due to the fact that there
are only two atom types and a very simple bonding mechanism involved. Clearly, when testing this approach with more demanding systems,
exhibiting covalent bonds or a larger number of atom types, considerably longer runs are to be expected. 

We generated different initial atom positions by using different start values
for the random number generator. About 70 runs were performed, of which
about half converged to
reasonable structure candidate, whereas the other half
remained in energetically unfavorable situations such as very low
densities or two-dimensional structures. This could certainly have been
improved by performing longer simulated annealing runs, but only at a much higher
computational cost.

After the quench was finished, the space group of the configuration was
analyzed with the program KPLOT \cite{Hundt79}, using algorithms
to identify the symmetry \cite{Hundt99a} and to find the space group
\cite{Hannemann98a}.
A subsequent local optimization was performed with
the CRYSTAL code, using analytical gradients for the nuclear positions
\cite{IJQC,CPC}
and the unit cell \cite{KlausDovesiRO,KlausDovesiRO1d2d} 
and the full geometry optimization as implemented
in the present release \cite{Mimmo2001,CRYSTAL2006}. 
As this local optimization is not too demanding in
terms of CPU time and as a high accuracy is desirable, the
integral thresholds and the threshold
for SCF convergence were set to the default 
values, and a shrinking factor of 4 was used for the $\vec k$ net. Also,
the original basis set as in \cite{Prencipe} was used.
The optimization was performed both at the Hartree-Fock level and at the level
of the local density approximation (LDA).
The fully optimized structures were again analyzed with KPLOT. 

The computational effort was typically a few days for the simulated annealing
and subsequent quench runs, and a few minutes up to one hour for the 
local optimization, on a single CPU of a standard PC.

\section{Results}
\label{res}

The results of these optimizations are displayed in Tables 1 and 2. Eight promising low-energy structure
candidates were found (shown in Figs.\ \ref{fig1} - \ref{fig3}, generated
with XCrysDen \cite{XCrysDen}): 
the rock salt structure as observed experimentally,
the zincblende structure (sphalerite), the wurtzite structure, the so-called
5-5 structure\cite{Schoen95,Schoen96b} (an ionic analogue to the $B_k$
structure of hexagonal BN), the NiAs structure and three
structures with space group 62, 7 and 36, denoted LiF(I), LiF(II) and
LiF(III), respectively (see Table 1 regarding the fraction of runs that
resulted in the various structure types). LiF(I) and LiF(II) consist of nets
of LiF$_4$-tetrahedra, with the first one containing narrow channels and the
second one resembling a twisted sphalerite or wurtzite structure. Finally,
LiF(III) consists of a network of LiF$_5$ square-pyramids. All these structure
candidates had also been observed in earlier global searches in alkali halide
systems\cite{Schoen95} using emipirical potentials consisting of a
Coulomb-term and a van-der-Waals-term, and one should note that LiF(I),
LiF(II) and LiF(III) are quite typical representatives of the higher-lying
local minima. This result clearly demonstrates,
that the global exploration of the energy landscape on the {\it ab initio}
level is feasible and provides reasonable structures.

\section{Discussion and Conclusion}
\label{disc}

Concerning the accuracy of the {\it ab initio} calculations, we note that at the Hartree-Fock level,
the wurtzite structure exhibits the lowest energy, in contrast to the experimental observation that LiF is found in the rock salt type. The failure
of the Hartree-Fock approach to account for the proper ground state
may be attributed to the neglect of the van der Waals interaction which
is important for alkali halides as was shown in \cite{Dollalkali1,Dollalkali2}.
Although the van der Waals interaction is not considered in the LDA either, the LDA performs better and predicts the rock salt
structure as the ground state, perhaps due to LDA's inherent tendency to
over-bind and favor higher coordinations. These observations are in good
agreement with {\it ab initio} calculations by {\v{C}}an{\v{c}}arevi{\'{c}} et
al\cite{Cancarevic06b,Chane06} using Hartree-Fock and density
functional theory, which showed that for LiF the functionals LDA
found the rocksalt type as the minimum structure, while Hartree-Fock
and the hybrid functional B3LYP found the wurtzite
type, respectively. Regarding the thermodynamic stability of the various modifications
as function of pressure, there is no change compared to the results of 
{\v{C}}an{\v{c}}arevi{\'{c}} et
al\cite{Cancarevic06b,Chane06}, which were based on the global exploration of the
empirical potential based enthalpy landscapes of LiF for ten different pressures ranging
from -16 GPa to +160 GPa. Here, we only give the transition pressures based on
the LDA functional: $\beta$-BeO $\rightarrow$ wurtzite at $\sim$ -8.5 GPa, wurtzite $\rightarrow$
5-5 at $\sim$ -5.5 GPa, and 5-5 $\rightarrow$ rock salt at $\sim$ -5.0 GPa. Possible high-pressure
phases might be the NiAs-type or the CsCl-type according to some density functionals; 
however, even if these modifications were to become thermodynamically stable, in both instances the transition pressures would be very high ($>$ 100 GPa) beyond the range of the validity of straightforward
{\it ab initio} calculations.

\begin{figure}
\includegraphics[width=5cm]{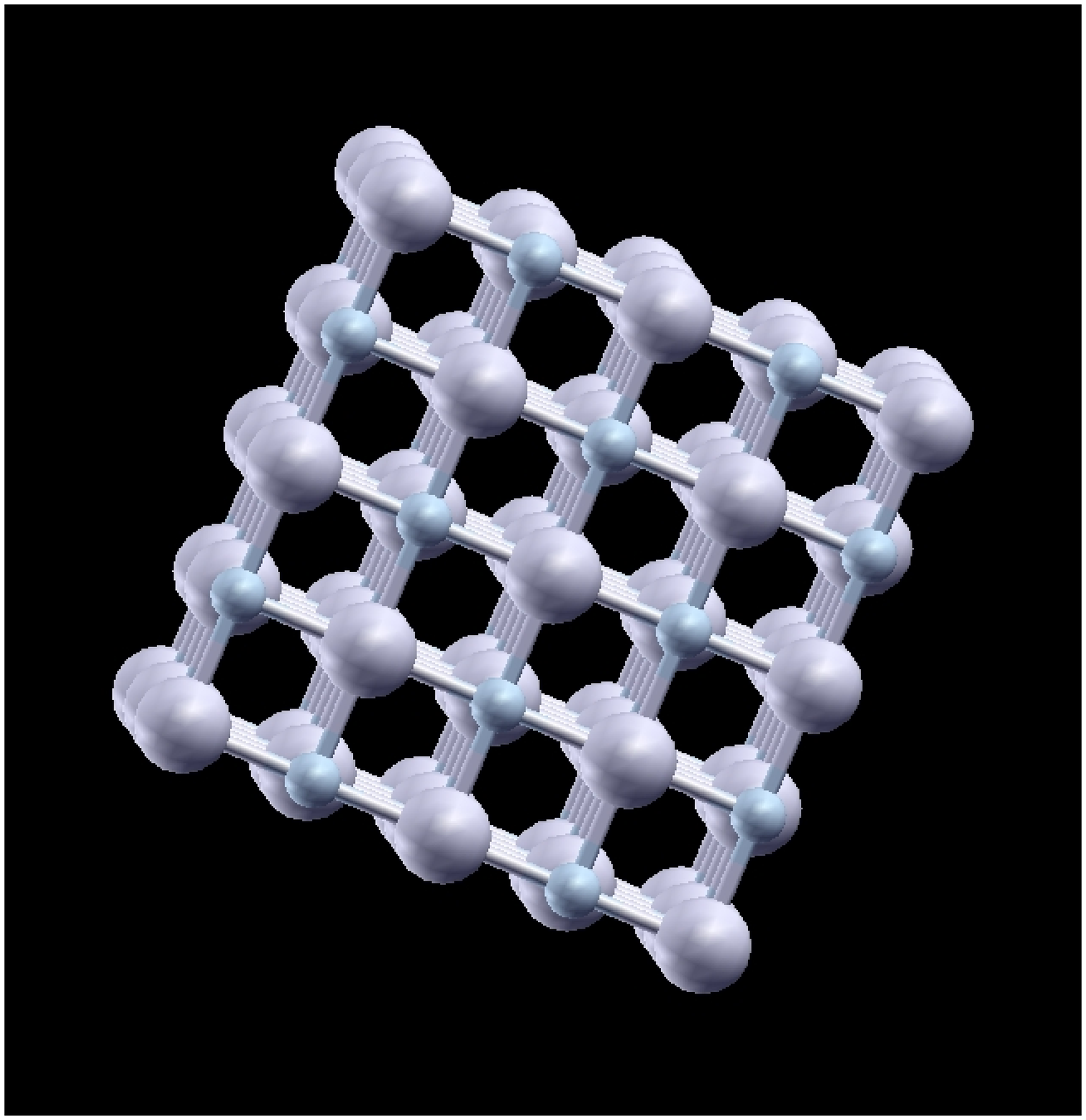} 
\includegraphics[width=5cm]{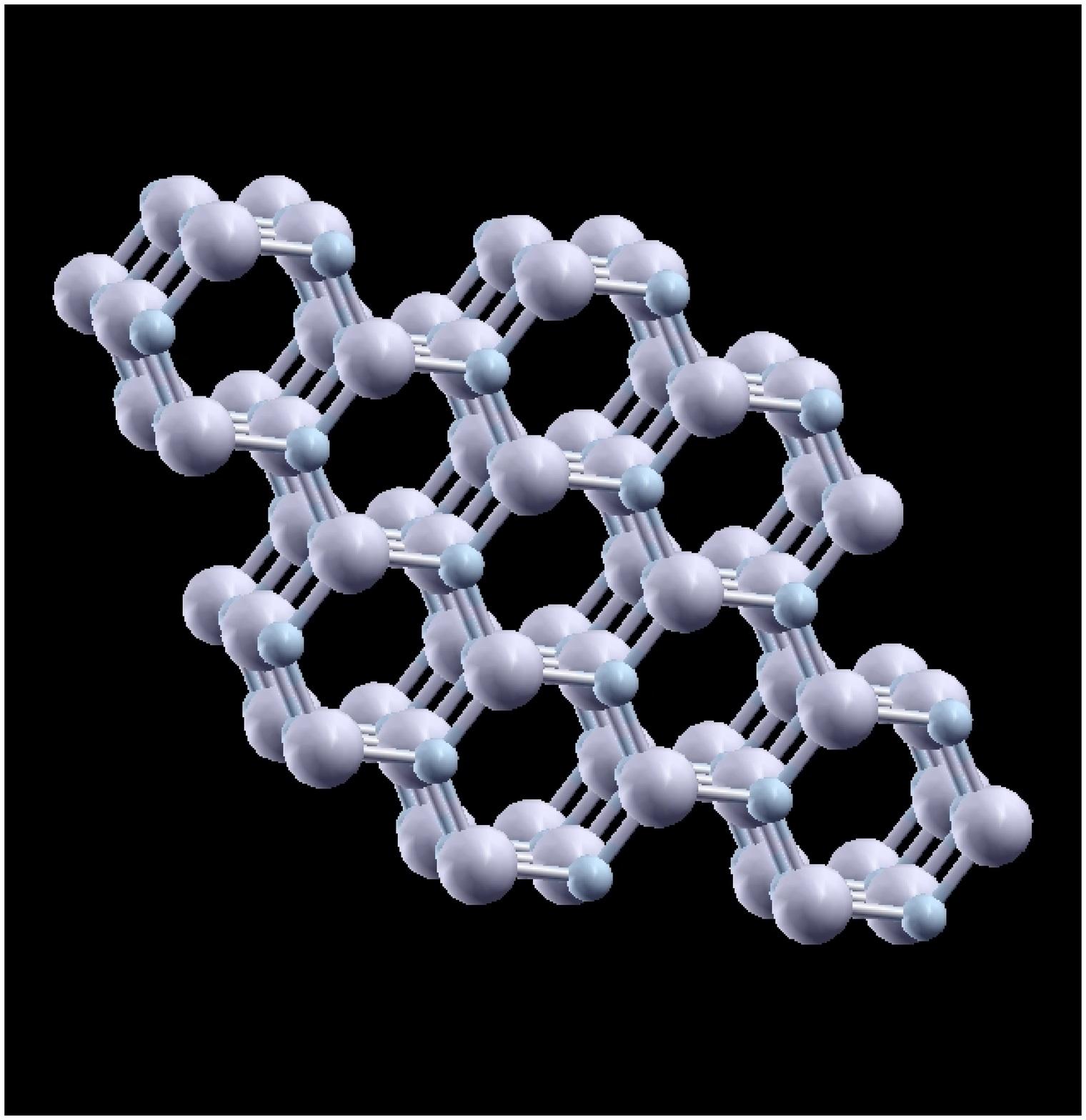} 
\includegraphics[width=5cm,angle=90]{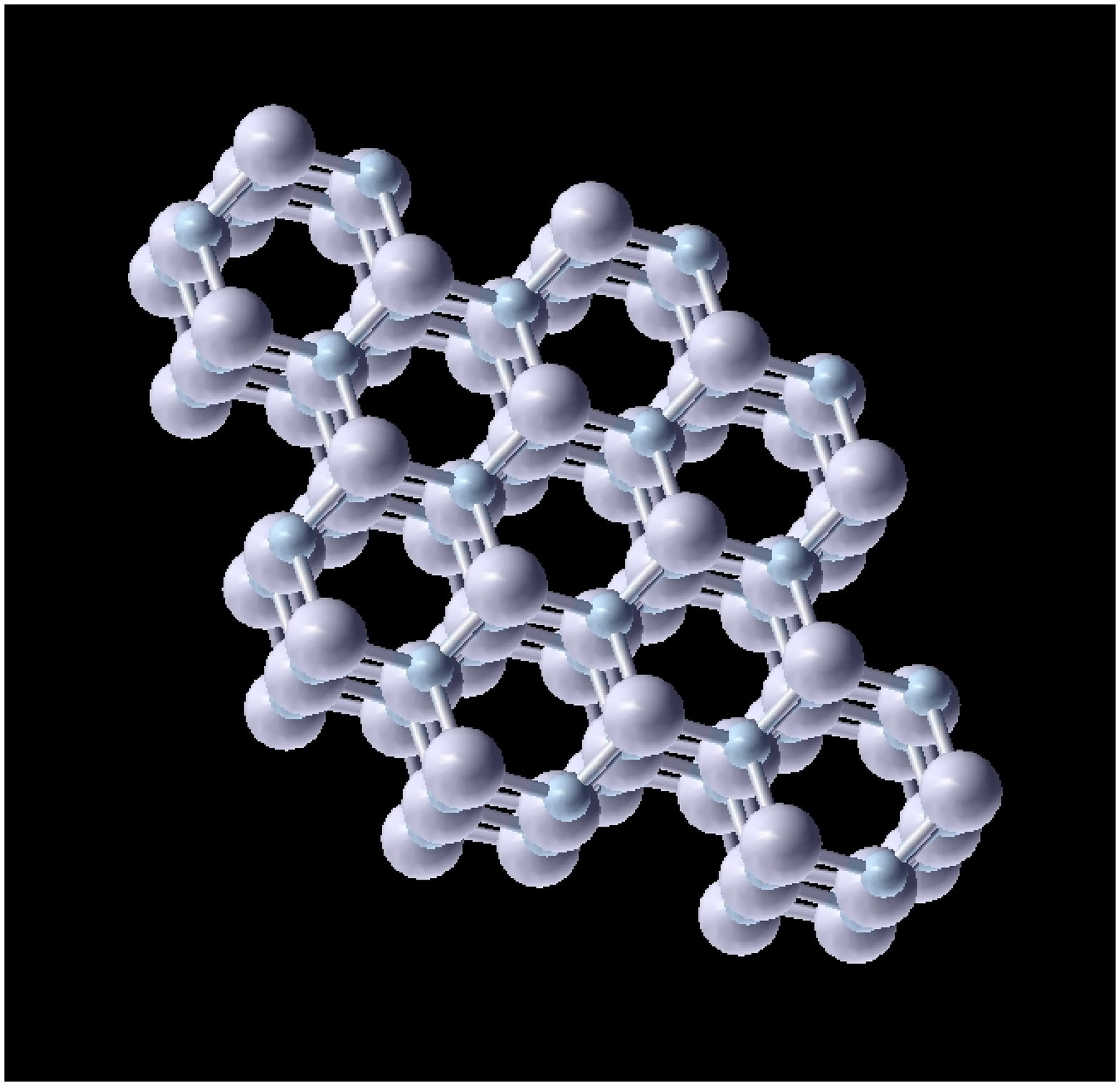} 
\caption{Ball-and-stick models of the structurally related rock salt (left), 5-5 (middle) and
wurtzite (right) structures, viewed along the c-axis. Large and small spheres correspond to Li- and F-atoms, respectively. The wurtzite-type transforms into the 5-5-type by slightly displacing the Li- and F-atoms
along the c-axis such that the local coordination changes from LiF$_4$-tetrahedra to LiF$_5$-trigonal bipyramids.
Similarly, one obtains the rock salt-type from the 5-5-type by compressing the 5-5-structure along
the a-axis such that the LiF$_5$-trigonal bipyramids become LiF$_6$-'square bipyramids', i.e.\ LiF$_6$-octahedra.}
\label{fig1}
\end{figure}

\begin{figure}
\includegraphics[width=5cm]{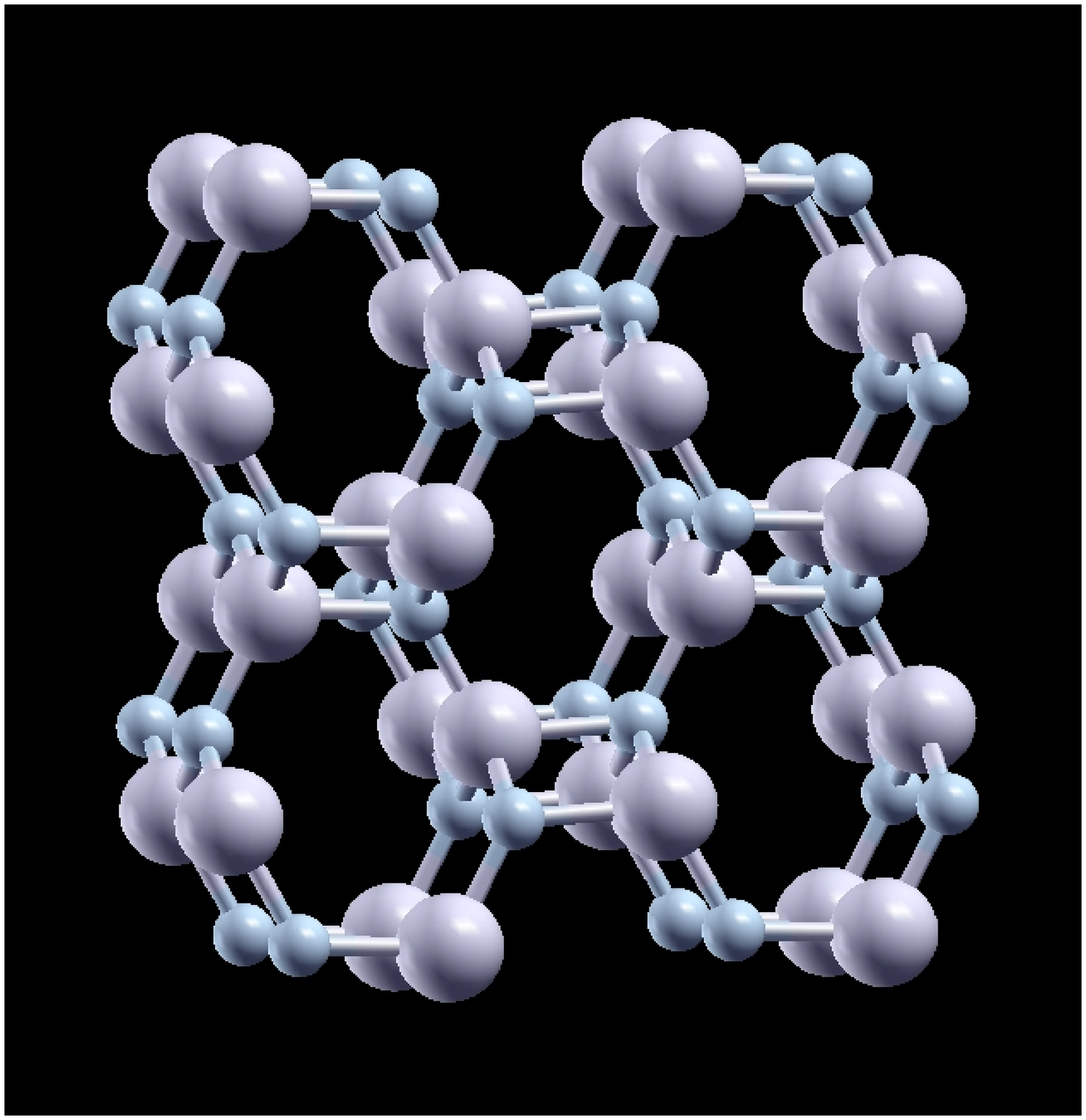} 
\includegraphics[width=5cm]{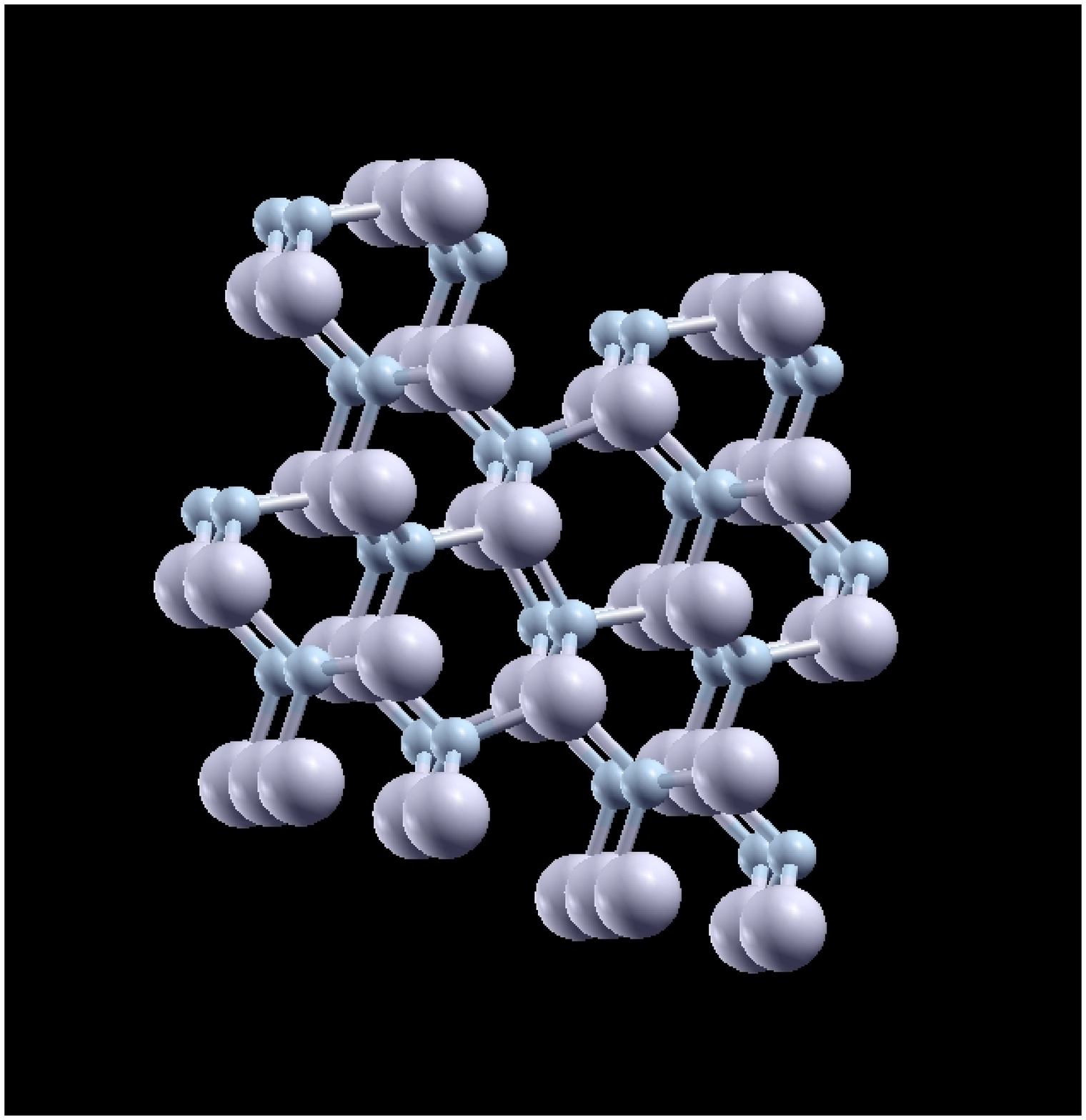} 
\includegraphics[width=5cm]{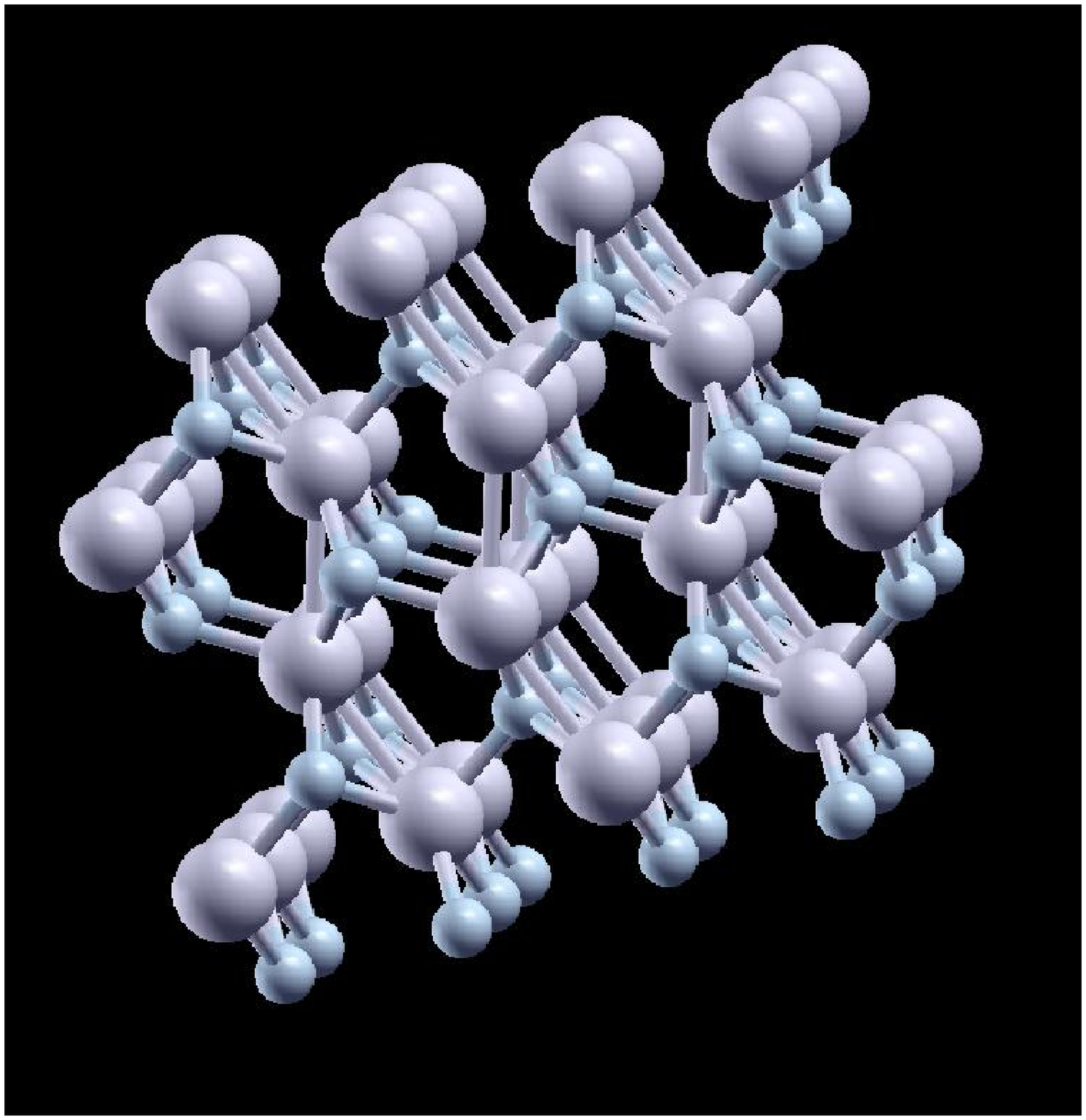} 
\caption{Ball-and-stick models of LiF(I) (left), LiF(II) (middle) and LiF(III) (right), respectively. For notation,
c.f.\ Fig.\ \ref{fig1}. These networks
of LiF$_4$-tetrahedra and LiF$_5$-square pyramids are characteristic of metastable higher-lying
local minima on the energy landscape of the alkali halides. \cite{Schoen95}}
\label{fig2}
\end{figure}

\begin{figure}
\includegraphics[width=5cm]{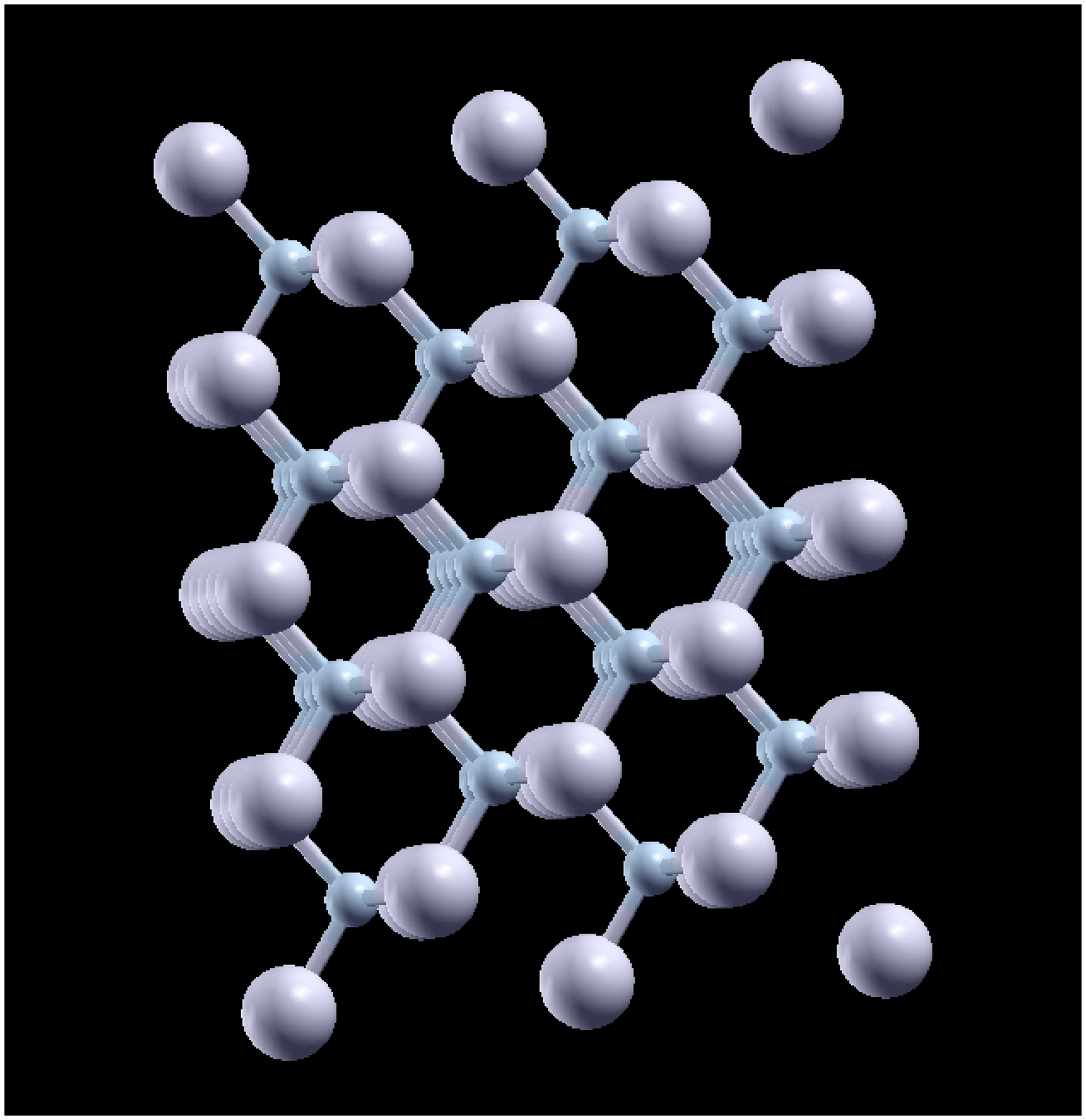} 
\includegraphics[width=5cm]{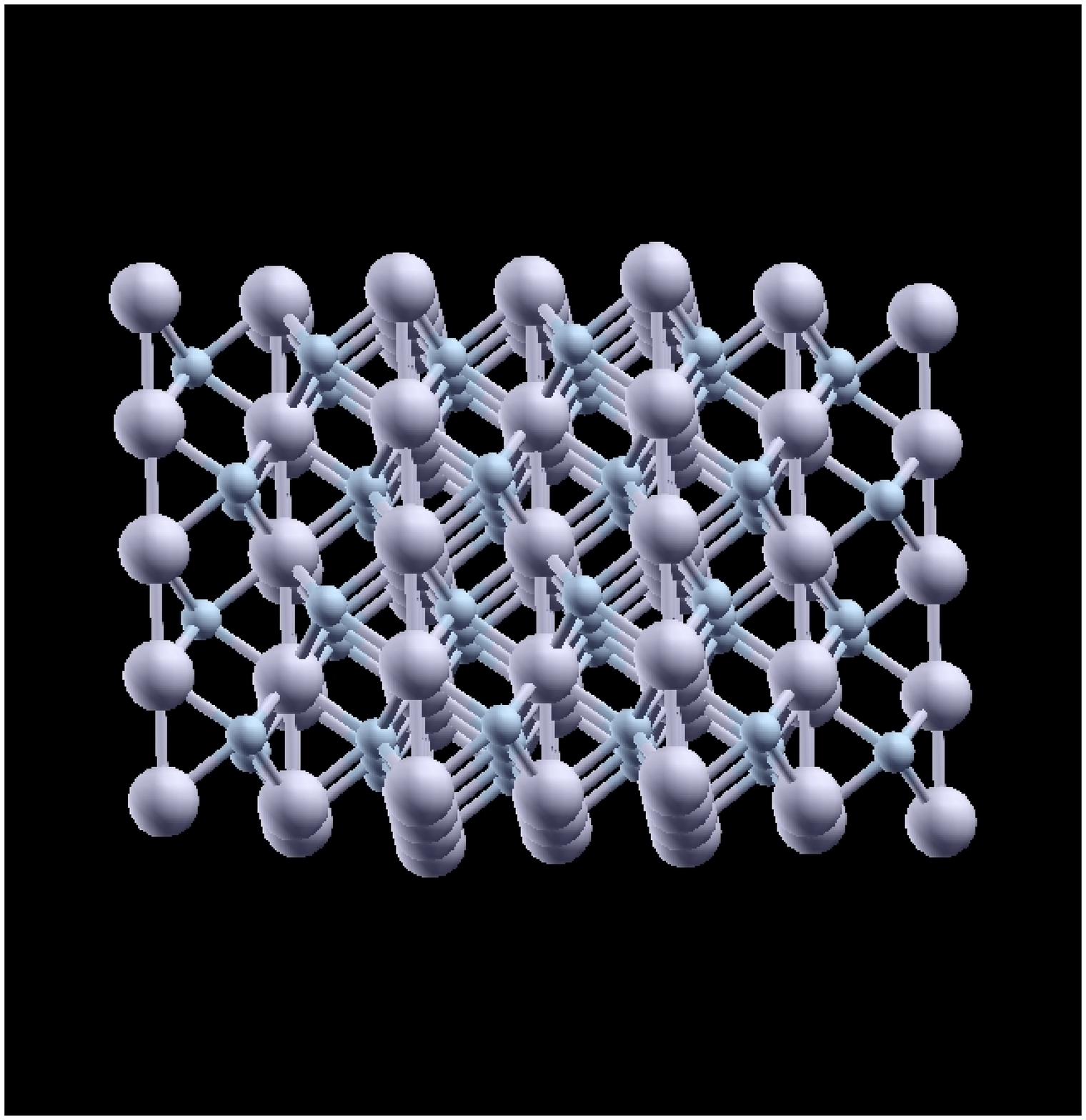} 
\caption{Ball-and-stick models of the sphalerite (left) and NiAs (right) structure, respectively. For notation,
c.f.\ Fig.\ \ref{fig1}.}
\label{fig3}
\end{figure}

The comparison between the outcome of the global landscape explorations
on the {\it ab initio} level and on the level of empirical potentials shows that the present
investigation determines essentially all the minima with the lowest energies
that had been found during the earlier work \cite{Schoen95}.
Due to the relatively short simulated annealing runs, at the end of the global exploration phase many of the candidates
had ended up in side-minima corresponding to distorted versions of the structures
belonging to the main minimum of the various basins typically exhibiting lower symmetries.
However, the subsequent local minimization using the standard high accuracy
for the {\it ab initio} energy calculations resulted in the system reaching the main
minima of the various basins. We note that in several 
instances the Hartree-Fock
and the DFT (LDA) calculations reached different structure types, wurtzite and
5-5, respectively. This supports our earlier observations that the wurtzite and
the 5-5-structure are close neighbors on the energy landscape.
\cite{footnote1}
Similarly, we observed that the
local mimization on DFT (LDA) level starting from four slightly distorted versions of
the LiF(I) structure type, resulted in one case in the rock salt and in the other
three cases in the 5-5-structure, respectively.
\cite{footnote2}
Again, this
confirms our earlier results that the 5-5-type is located on the landscape 
close
to the rock salt structure, possibly constituting a transitional modification on the route
from the wurtzite to the rock salt type (c.f.\ Fig.\ \ref{fig1}).
\cite{footnote3}
Finally, we note that
in contrast to the study using the empirical energy landscape the structures exhibiting
four- and five-fold coordination were found quite often. This is clearly a
reflection of the fact that their energies on the {\it ab initio} level are very similar to the ones
of the structure types with six-fold coordination (rock salt, NiAs) which are more strongly
preferred when using the empirical potential.

To summarize, the comparison between the empirical and {\it ab initio} energy
landscapes shows that at least for the case of a simple ionic system such as
LiF a) the minima representing the most relevant modifications and similarly
most of the other chemically interesting structure types are present on both
energy landscapes, b) however, their ranking in energy depends on the type of
energy calculation and similarly the likelihood of observing the minima can
also be quite different, c) structures which are closely related on the
empirical landscape, i.e. separated by relatively small barriers, are also
close neighbors on the {\it ab initio} energy landscape, and d) even classes of structures that are rather unusual such as those containing channels or square pyramids are found on both landscapes. We can thus conclude that one of the fundamental
assumptions behind the standard approach of structure prediction of solids, i.e.\ that the empirical
potentials can be employed in global landscape explorations to identify the relevant modifications
of a solid compound is valid at least for ionic systems where reasonable suitable potentials are 
available.

Of course, the number of
simulated annealing runs possible when using the Hartree-Fock energy
function is still much smaller than the number of runs with an empirical
energy function. As a consequence, in particular the many high-lying
and/or shallow minima associated with structures containing channels or 
belonging to transitions between two large basins, were detected relatively rarely.
Nevertheless, the fact that all relevant modifications have been observed
in the present study shows that the global exploration of the energy
landscape of solids on the {\it ab initio} level using standard simulated
annealing as the global optimization tool has finally become feasible. This will
allow us to predict the possible modifications of crystalline compounds also
in those chemical systems, where no simple empirical potentials can be
constructed, thus overcoming one of the major hurdles facing crystal structure
prediction in general chemical systems.

\acknowledgments
We would like to thank Z. {\v{C}}an{\v{c}}arevi{\'{c}} 
and U. Wedig for valuable discussions.
The work was funded by the MMM-initiative of the Max-Planck-Society.


\begin{table}[ht]
\begin{center}
\caption{\label{energy} Structures obtained corresponding to local minima,
the total energy (without a correction for the
basis set superposition error)
for four formula units, in hartree units ($E_h$),
and the number of times the structures were found.}
\begin{tabular}{cccccc}
\hline 
structure type & space group & \multicolumn{2}{c}{energy ($E_h$)} & 
\multicolumn{2}{c}{number of times found}\\ 
& & HF & LDA & HF & LDA\\
rock salt & 225 & -428.2210 & -427.0665 & 9 & 11 \\ 
zincblende & 216 & -428.2211 & -427.0445 & 10 & 9 \\ 
5-5 & 194 & -428.2222 & -427.0534 & 4 & 12 \\
wurtzite & 186 & -428.2250 & -427.0484 & 8 & 4 \\ 
NiAs & 194 & -428.2051 & -427.0515 & 1 & 1 \\ 
LiF(I) & 62 & -428.2162 & - & 4 & 0\\ 
LiF(II) & 7 & -428.2089 & -427.0374 & 1 & 1 \\ 
LiF(III) & 36 & -428.2054 & - & 1 & 0 \\
\end{tabular}
\end{center}
\end{table}

\begin{table}[ht]
\caption{\label{geometry}
Optimized geometry of the various structures obtained. \\
$^a$Cell parameter from
experiment (rock salt-type structure): $a$=4.027 \AA \cite{Streltsov87}}
\begin{tabular}{ccc}
\hline 
structure type & \multicolumn{2}{c}{lattice constant, angle and Wyckoff
  positions} \\
 (space group) &  \multicolumn{1}{c}{HF} &  \multicolumn{1}{c}{LDA} \\ \\
rock salt$^a$ (225) &  $a$=4.01 \AA & $a$=3.94 \AA \\
&  Li (0,0,0) &(0,0,0) \\
&  F  (1/2,0,0) &  (1/2,0,0) \\ \\
zincblende (216) & $a$=4.31 \AA & $a$=4.23 \AA \\
& Li  (0,0,0) & (0,0,0) \\
& F  (1/4, 1/4, 1/4) 
& (1/4, 1/4, 1/4) \\ \\
5-5 (194) & $a$=3.28 \AA; $c$=4.05 \AA & $a$=3.24 \AA; $c$=3.94 \AA \\
& Li (1/3,2/3,1/4) & (1/3,2/3,1/4) \\
& F (2/3,1/3,1/4) &  (2/3,1/3,1/4)\\ \\
wurtzite (186) & $a$=3.09 \AA; $c$=4.86 \AA & $a$=3.07 \AA; $c$=4.64 \AA \\
& Li (1/3,2/3,0) & (1/3,2/3,0) \\
& F (1/3,2/3,0.386) &  (1/3,2/3,0.399)\\
 \\
NiAs (194) & $a$=2.79 \AA; $c$=4.80 \AA & $a$=2.73 \AA; $c$=4.74 \AA \\
& Li (0,0,0) & (0,0,0) \\
& F (1/3,2/3,1/4) &
(1/3,2/3,1/4) \\ \\
LiF(I) (62) & $a$=5.60 \AA, $b$=3.14 \AA, $c$=5.17 \AA & \\
& Li (0.831, 3/4, 0.587) &  - \\
& F (0.677, 1/4, 0.895) &  \\ \\
LiF(II) (7)& $a$=2.92 \AA, $b$=5.62 \AA, $c$=5.34 \AA, $\beta=$115.9$^\circ$ 
& $a$=2.84 \AA, $b$=5.50 \AA, $c$=5.25 \AA; $\beta$=115.5$^\circ$ \\
& Li (0, 0.407,0) & (0, 0.405,0) \\
& Li (0.648, 0.121, 0.428) & (0.638, 0.122, 0.430) \\
& F (0.955, 0.880, 0.676) & (0.941, 0.878, 0.678) \\
& F (0.613, 0.403, 0.616) & (0.599, 0.401, 0.617) \\ \\
LiF(III) (36) & $a$=2.72 \AA, $b$=5.30 \AA, $c$=4.81 \AA &  \\
& Li (0, 0.083, 0) &  - \\
& F (0, 0.617,0.714) & \\
\end{tabular}
\end{table}

\clearpage







\begin{thebibliography}{99}
\bibitem{footnote1}
{Note that the wurtzite structure has a lower energy than the 
5-5-structure at the Hartree-Fock level, and
that the reverse holds true at the DFT (LDA) level.}
\bibitem{footnote3}
{This is supported by a study on the stability of MgO in the 
wurtzite-type, where the 5-5 structure was rediscovered and called h-MgO by
the authors\cite{Lambrecht01}.}
\end{thebibliography}
\end{document}